\documentclass[journal=rsm]{CUP-JNL-DTM}%

\usepackage{graphicx}
\usepackage{multicol,multirow}
\usepackage{amsmath,amssymb,amsfonts}
\usepackage{mathrsfs}
\usepackage{amsthm}
\usepackage{rotating}
\usepackage{appendix}
\usepackage{ifpdf}
\usepackage[T1]{fontenc}
\usepackage{newtxtext}
\usepackage{newtxmath}
\usepackage{textcomp}
\usepackage{xcolor}
\usepackage{lipsum}
\usepackage[colorlinks,allcolors=blue]{hyperref}

\theoremstyle{definition}

\numberwithin{equation}{section}

\jyear{2025}

\begin{document}

\newcommand{\sumr}{\sum_{s = 1}^{G-1}}
\newcommand{\sumk}{\sum_{r = 1}^s}
\newcommand{\psr}{p_{gr}}
\newcommand{\sucra}{\text{SUCRA}}
\newcommand{\E}{\text{E}}
\newcommand{\MVN}{\text{MVN}}

\newcommand{\Yij}{Y_{ij}}
\newcommand{\Xij}{\bm{X_{ij}}}
\newcommand{\Aij}{\bm{A_{ij}}}
\newcommand{\aij}{\bm{a_{ij}}}
\newcommand{\ti}[1]{t_{i#1}}
\newcommand{\bbetai}{\bm{\beta_{i}}}
\newcommand{\xij}{\bm{x_{ij}}}
\newcommand{\aijk}{a_{ij\ti{k}}}
\newcommand{\xijb}{\bm{x_{ij}^{(\beta)}}}
\newcommand{\xijp}{\bm{x_{ij}^{(\delta)}}}
\newcommand{\xp}{x_{1}, \dots, x_Q}
\newcommand{\xx}{\bm{x}}

\newcommand{\psibarl}[2]{\psi_{#1}^{(#2)}}
\newcommand{\psil}[2]{\psi_{#1}^{(#2)}}
\newcommand{\psilj}[1]{\psi_{#1}}

\newcommand{\deltaim}{\underline{\deltai}}
\newcommand{\psim}{\underline{\psivec}}
\newcommand{\Vi}{\bm{V_i}}

\newcommand{\deltai}{\bm{\delta_i}^{(1)}}
\newcommand{\deltaa}{\bm{\delta}^{(1)}}
\newcommand{\deltaihat}{\bm{\hat{\delta_i}}^{(1)}}
\newcommand{\Si}{\bm{S_i}}

\newcommand{\psivec}{\bm{\psi}^{(1)}}

\newcommand{\deltail}[1]{\delta_{i,#1}^{(1)}}
\newcommand{\deltailhat}[1]{\hat{\delta}_{i,#1}^{(1)}}

\begin{Frontmatter}

\title[Personalized Treatment Hierarchies in Bayesian NMA]{Personalized Treatment Hierarchies in Bayesian Network Meta-Analysis}

\author[1]{Augustine Wigle}
\author[1]{Erica E. M. Moodie}

\authormark{Augustine Wigle \& Erica Moodie}

\address[1]{\orgdiv{Department of Epidemiology, Biostatistics and Occupational Health}, \orgname{McGill University}, \orgaddress{\street{2001 McGill College Ave}, \postcode{H3A 1G1}, \state{Quebec}, \country{Canada}}}



\abstract{Network Meta-Analysis (NMA) is an increasingly popular evidence synthesis tool that can provide a ranking of competing treatments, also known as a treatment hierarchy. Treatment-Covariate Interactions (TCIs) can be included in NMA models to allow relative treatment effects to vary with covariate values. We show that in an NMA model that includes TCIs, treatment hierarchies should be created with a particular covariate profile in mind. We outline the typical approach for creating a treatment hierarchy in standard Bayesian NMA and show how a treatment hierarchy for a particular covariate profile can be created from an NMA model that estimates TCIs. We demonstrate our methods using a real network of studies for treatments of major depressive disorder.}

\end{Frontmatter}

\section*{Highlights}
What is already known: 
\begin{itemize}
    \item Network meta-analyses (NMAs) that include treatment-covariate interactions, such as individual participant data (IPD) NMAs, provide relative treatment effects conditional on covariate values.
    \item Treatment hierarchies created from ranking metrics are a popular way to summarise the results of a network meta-analysis.
\end{itemize}

What is new:
\begin{itemize}
    \item We describe how existing ranking metrics can be easily adapted to provide personalized treatment hierarchies from an IPD NMA.
\end{itemize}

Potential impact for Research Synthesis Methods readers:
\begin{itemize}
    \item Practitioners should consider creating tools that can compute ranking metrics in IPD NMAs to support personalized clinical decision-making.
\end{itemize}

\section{Introduction}

Network meta-analysis (NMA) is a statistical methodology that facilitates the estimation of relative effects for multiple treatments from a set of clinical trials. A convenient method for summarizing the results of a standard NMA is to construct a treatment hierarchy, that is, a list of the treatments ranked in order of preference. Treatment hierarchies are constructed using ranking metrics which consider the magnitude and uncertainty of relative effect estimates, such as the Surface Under the Cumulative RAnking curve (SUCRA) in Bayesian NMA \cite{salanti_introducing_2022, salanti_graphical_2011}. A treatment hierarchy is a desired outputs of many NMAs \cite{papakonstantinou_answering_2022}.

When effect modifiers are present, incorporating Treatment-Covariate Interactions (TCIs) into the NMA model can reduce heterogeneity and provide more precise estimates \cite{donegan_combining_2013-1,freeman_framework_2018, riley_using_2023}. TCIs allow the expected relative treatment effects, and therefore treatment hierarchies, to differ between individuals with different covariate values, presenting an opportunity to compute personalized hierarchies. Authors have called for the reporting of treatment hierarchies at different covariate levels when TCIs are included in an NMA \cite{donegan_combining_2013-1,freeman_framework_2018, riley_individual_2020, riley_using_2023}, but this is rarely done in practice. In this article, we show how Bayesian NMAs with TCIs can produce personalized treatment hierarchies, and demonstrate the methods using a network of studies for Major Depressive Disorder (MDD).

\section{Treatment Hierarchies in Standard NMA Models}

In a standard NMA with no TCIs, the parameters representing treatment performance are given by $\psil{20}{1}, \psil{30}{1}, \dots, \psil{G0}{1}$, where $\psil{g0}{1}$ is the expected relative effect of treatment $g$ compared to treatment 1, and the index $0$ indicates that the parameter represents the main effect of the treatment and not an interaction, which will become important in the following section. Additionally, $G$ is the number of treatments in the network, and $\psil{10}{1} = 0$. We assume that the expected relative effect of treatment $g$ versus $h$, $\psil{g0}{h}$, is given by $\psil{g0}{1}-\psil{h0}{1}$ for any $g,h$ in $1, \dots, G$. This is referred to as the consistency assumption in NMA literature \cite{ades_twenty_2024}. In the remainder of the article, we suppress the superscript $(1)$ and emphasize that all parameters are relative to treatment 1.

SUCRA values are calculated from the relative effects to obtain a treatment hierarchy that is relevant to the population represented by the NMA. The SUCRA value for treatment $g$ is calculated in a standard Bayesian NMA as
\begin{align} 
    \sucra(g) &= \frac{\sumr \sumk \psr}{G-1} \notag \\
    &= \frac{G - \E\left[rank(g)\right]}{G-1}, \label{eq:sucra1}
\end{align}
where $\psr$ is the probability that treatment $g$ takes rank $r$, i.e., it has a more favorable response than exactly $r$ treatments, and $\E\left[rank(g)\right]$ is the average rank of treatment $g$ \cite{salanti_graphical_2011, rucker_ranking_2015}. The ranking probabilities or average ranks are estimated using samples from the posterior distribution of the relative effects, $\psilj{10}, \psilj{20}, \dots, \psilj{G0}$, as described in Supplementary Information (SI) \S 1.1 
\cite{rosenberger_predictive_2021}. The SUCRA of treatment $g$ in a standard NMA is interpreted as the proportion of competing treatments that $g$ is expected to beat \cite{salanti_graphical_2011}, or the inverse-scaled average rank of treatment $g$ \cite{rucker_ranking_2015}, \textit{when it is employed in the population represented by the NMA}. The resulting treatment hierarchy is not personalized to any particular patient and depends on the covariate distribution in the population \cite{phillippo_multilevel_2020}.

\section{Treatment Hierarchies in NMA Models with TCIs}

TCIs can be estimated using aggregate data NMA models with meta-regression and Individual Participant Data (IPD) NMA models, including recently proposed Individualized Treatment Rule (ITR) NMA models. In SI \S 2, we describe the suitability of each model type for creating personalized treatment hierarchies. Briefly, we do not recommend using aggregate data NMA meta-regression to create personalized treatment hierarchies due to its risk of ecological bias and confounding \cite{donegan_assessing_2012, riley_meta-analysis_2008, berlin_individual_2002, schmid_meta-regression_2004}. IPD NMA can be used to produce personalized treatment hierarchies when care is taken in specifying the model \cite{hua_one-stage_2017,riley_individual_2020,riley_using_2023}.

In an NMA model which includes TCIs, the interpretation of the parameter $\psilj{g0}$ does not correspond to the expected relative effect on average in the population. To illustrate, suppose there are $Q$ effect-modifying covariates included as TCIs in an NMA model. The expected relative effect of treatment $g$ versus treatment 1 for an individual with covariate values $X_1 = x_1, X_2 = x_2, \dots, X_Q = x_Q$ is given by
\begin{equation}\label{eq:exprel}
    \psilj{g0} + \psilj{g1} x_1 + \psilj{g2} x_2 + \dots + \psilj{gQ} x_Q,
\end{equation}
where $\psilj{g1}, \dots, \psilj{gQ}$ are the coefficients representing the interaction between each covariate and treatment $g$ compared to treatment 1, that is, $\psilj{gq}, q > 0$ is the change in the relative effect of treatment $g$ compared to treatment 1 for a one unit increase in covariate $q$. Therefore, $\psilj{g0}$ corresponds to the expected relative effect of $g$ versus 1 \textit{when all covariate values are equal to zero}.

Now, consider creating a treatment hierarchy from a model which includes TCIs. If we were to use only $\psilj{g0}$, $g = 1, \dots, G$ to create the hierarchy for such a model, the hierarchy would apply to patients whose covariate values are all equal to zero. The hierarchy will be meaningless for patients with a different covariate profile. Clearly, for a treatment hierarchy to be meaningful in an NMA with TCIs, it should be intentionally created for a particular set of covariate values.

Let $\E\left[rank(g)\mid \xp\right]$ be the average rank of treatment $g$ given covariates $\xp$. The average ranks are determined using the expected relative effects defined in equation \eqref{eq:exprel}. They can be calculated using samples from the posterior of $\psilj{gq}$, $g = 1,$ $\dots$, $G$ and $q = 0$, $\dots$, $Q$ and used to compute SUCRA values as described in SI \S 1.2. The treatment hierarchy produced in this way is personalized to a patient with the covariate profile defined by $\xp$.

\section{Example - Pharmacological Treatments for MDD}

We illustrate the use and impact of personalized treatment hierarchies using data from three studies investigating treatment for MDD. Details on the study designs are given in SI \S 3. The data provide information on six pharmacological treatments: Bupropion, Citalopram + Bupropion, Citalopram + Buspirone, Escitalopram, Sertraline, and Venlafaxine. The response is defined as the negative of the 17-question Hamilton Rating Scale for Depression (HRSD-17), so that larger values are preferred. The following covariates, considered to be potential effect modifiers for depression, were included as TCIs in the analysis: age, sex, marital status, years of education, employment status, household size, age of depression onset, number of depressive episodes, chronicity of current MDD episode, and HRSD-17 before receiving any treatment in the study. We fit a two-stage common effect Bayesian ITR NMA model with IPD to estimate expected relative treatment effects in the MDD network \cite{shen_two-stage_2025-1}, described in detail in SI \S 4. 

We computed personalized treatment hierarchies using SUCRA for two hypothetical patients, Patient A and Patient B, each with unique covariate profiles: Patient A is employed while Patient B is unemployed, Patient A has experienced three or fewer depressive episodes while Patient B has experienced more than three, Patient A is male and Patient B is not, and Patient A lives in a household with two more members than Patient B. The specific covariate values for each patient are described in SI \S 4.3.

\subsection{Results}

A plot showing point estimates and 95\% credible intervals for treatment effects and TCIs is shown in SI \S 4.4. All credible intervals contain zero. This is not too surprising, since only three studies were included in the analysis. In Figure \ref{fig:posterior}, we show the posterior distribution of the expected relative effects as defined in equation \eqref{eq:exprel} of each treatment compared to Sertraline for Patients A and B. It is clear that there are differences in the posterior densities between the patients. The SUCRA values for Patients A and B reflect the differences in the posteriors, and are compared in Table \ref{tab:rankings}. For example, Citalopram + Bupropion is ranked first for Patient A, while it is ranked only fourth out of the six treatments for Patient B.

\section{Discussion}

Personalized treatment hierarchies can be computed from IPD NMA models that include TCIs. We demonstrated the methods using an ITR NMA of treatments for MDD. Although none of the TCIs in the example were statistically significant (credible intervals contained zero), the posterior distributions of the expected relative effects changed for different patient profiles and resulted in different treatment hierarchies, suggesting possible value in personalizing recommendations. We emphasize that although individual interaction terms may not be statistically significant, their combined effects may still have an important influence on personalized ranking metrics. 

We focused on creating treatment hierarchies from the expected relative effects. In random-effects NMA models, treatment effects and interactions are allowed to vary across studies, and a personalized treatment hierarchy that better reflects uncertainty could be obtained by considering this additional predictive variability. A Bayesian approach to creating predictive treatment hierarchies in standard NMAs has been proposed \cite{rosenberger_predictive_2021} and could be adapted to personalized treatment hierarchies.

We echo previous calls to report treatment hierarchies for different covariate profiles when possible \cite{donegan_combining_2013-1, freeman_framework_2018, riley_individual_2020, riley_using_2023}. If many covariates are effect modifiers, it may not be feasible to report hierarchies for all covariate patterns of interest. Some recent IPD NMAs have included online tools that allow for individuals to compute expected relative effects for custom covariate values to aid in decision-making, e.g. \cite{karyotaki_internet-based_2021, furukawa_dismantling_2021}. With little additional effort, personalized treatment hierarchies could be included in the output of such tools. We encourage researchers conducting IPD NMAs to consider creating tools that can provide personalized treatment hierarchies to support clinical decision-making.

\begin{Backmatter}


\paragraph{Funding Statement}
AW is supported by Natural Sciences and Engineering Research Council of Canada (grant number PDF-598932-2025). EEMM is a Canadian Institutes of Health Research Canada Research Chair (Tier 1) in Statistical Methods for Precision Medicine.

\paragraph{Competing Interests}
None.

\paragraph{Data Availability Statement}
Code and data to reproduce the results in this article are available on GitHub at \url{https://github.com/augustinewigle/personalized_trt_hierarchies}.

\paragraph{Ethical Standards}
The research meets all ethical guidelines, including adherence to the legal requirements of the study country.

\paragraph{Author Contributions}
\textbf{AW} - Conceptualization, methodology, software, formal analysis, data curation, writing - original draft, writing - review \& editing, visualization.
\textbf{EEMM} - Conceptualization, supervision, writing - review \& editing.


\begin{figure}[h]
    \centering
    \includegraphics[width=\linewidth]{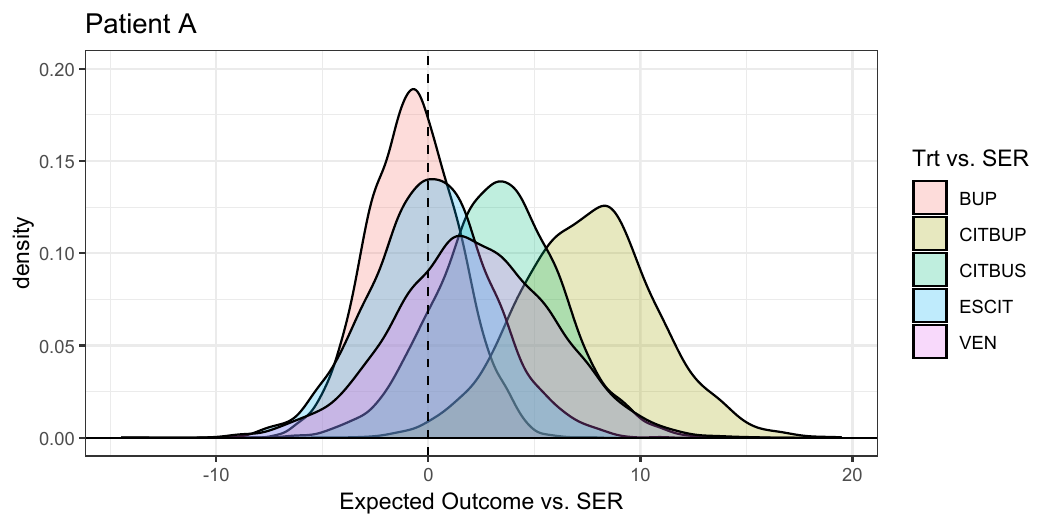}
    \includegraphics[width=\linewidth]{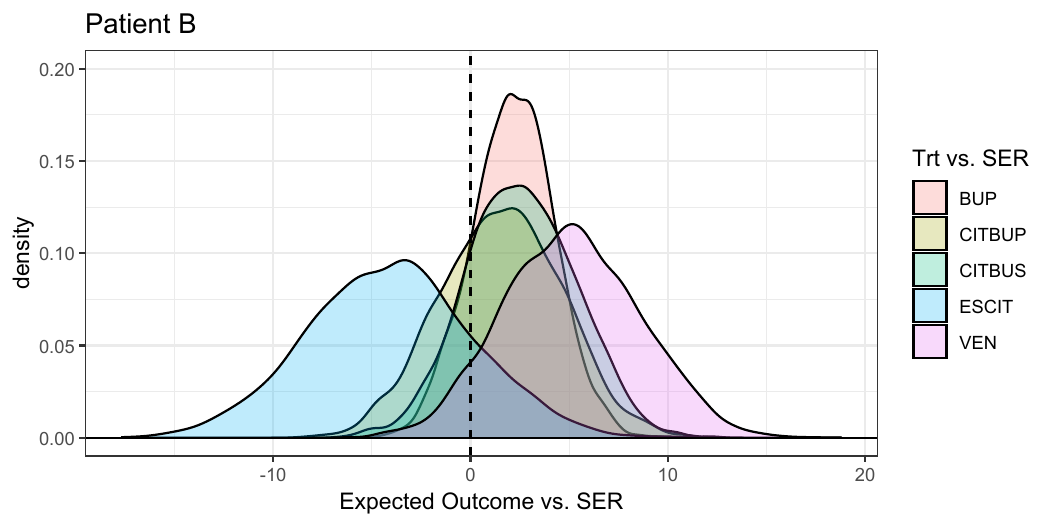}
    \caption{Posterior distributions of the relative effects of each treatment for Patients A and B}
    \label{fig:posterior}
\end{figure}

\begin{table}[ht]
    \caption{SUCRA values and ranking of each treatment for Patient A and Patient B}
    \label{tab:rankings}
    \begin{tabular}{l|cc|cc}
        \multirow{2}{*}{Treatment} & \multicolumn{2}{c|}{Patient A} & \multicolumn{2}{c}{Patient B}\\
        & SUCRA & Rank & SUCRA &  Rank \\ \hline
        Bupropion & 0.21 & 6 & 0.60 & 3 \\ 
        Citalopram + Bupropion & 0.96 & 1 & 0.50 & 4 \\ 
        Citalopram + Buspirone & 0.66 & 2 & 0.62 & 2 \\ 
        Escitalopram & 0.34 & 4 & 0.10 & 6 \\ 
        Venlafaxine & 0.52 & 3 & 0.86 & 1 \\ 
        Sertraline & 0.31 & 5 & 0.31 & 5 \\ \hline
    \end{tabular}
    \footnotetext{SUCRA: Surface Under the Cumulative RAnking curve.}
\end{table}


\printbibliography

\end{Backmatter}

\end{document}